\documentclass[aip,jap,preprint]{revtex4-1}

\usepackage{amsmath}
\usepackage{amssymb}
\usepackage{graphicx}
\usepackage{dcolumn}
\usepackage{bm}
\usepackage{physics}
\usepackage{caption}
\usepackage{subcaption}
\usepackage{booktabs}
\usepackage{multirow}
\usepackage{color}
\usepackage{soul}
\usepackage{epstopdf}
\newcommand{\sign}{\text{sgn}}
\usepackage{comment}
\usepackage[]{lineno}

\begin{document}


\title{Skyrmion-induced bound states on the surface of 3D Topological Insulators} 



\author{Dimitrios Andrikopoulos}
\email[]{dimitrios.andrikopoulos@imec.be}
\affiliation{KU Leuven, ESAT, Kasteelpark Arenberg 10 , Leuven, B-3001 Belgium}
\affiliation{imec, Kapeeldreef 75, Leuven, 3001 Belgium}

\author{Bart Sor\'{e}e}%
\email[]{bart.soree@imec.be}
\affiliation{KU Leuven, ESAT, Kasteelpark Arenberg 10 , Leuven, B-3001 Belgium}
\affiliation{Universiteit Antwerpen, Physics Dpt., Condensed Matter Theory, Groenenborgerlaan 171, Antwerpen, B-2020 Belgium}
\affiliation{imec, Kapeeldreef 75, Leuven, 3001 Belgium}

\author{Jo De Boeck}
\affiliation{KU Leuven, ESAT, Kasteelpark Arenberg 10 , Leuven, B-3001 Belgium}
\affiliation{imec, Kapeeldreef 75, Leuven, 3001 Belgium}


\date{\today}

\begin{abstract}
The interaction between the surface of a 3D topological insulator and a skyrmion / anti-skyrmion structure is studied in order to investigate the possibility of electron confinement due to skyrmion presence. Both hedgehog (N\'{e}el) and vortex (Bloch) skyrmions are considered. For the hedgehog skyrmion the in-plane components cannot be disregarded and their interaction with the surface state of the TI has to be taken into account. A semi-classical description of the skyrmion chiral angle is obtained using the variational principle. It is shown that both the hedgehog and the vortex skyrmion can induce bound states on the surface of the TI. However, the number and the properties of these states depend strongly on the skyrmion type and on the skyrmion topological number $N_{Sk}$. The probability densities of the bound electrons are also derived where it is shown that they are localized within the skyrmion region.
\end{abstract}


\maketitle 


\section{\label{sec:intro}Introduction}
\begin{figure*}
\centering
\begin{subfigure}{.5\textwidth}
  \centering
  \includegraphics[width=.55\linewidth]{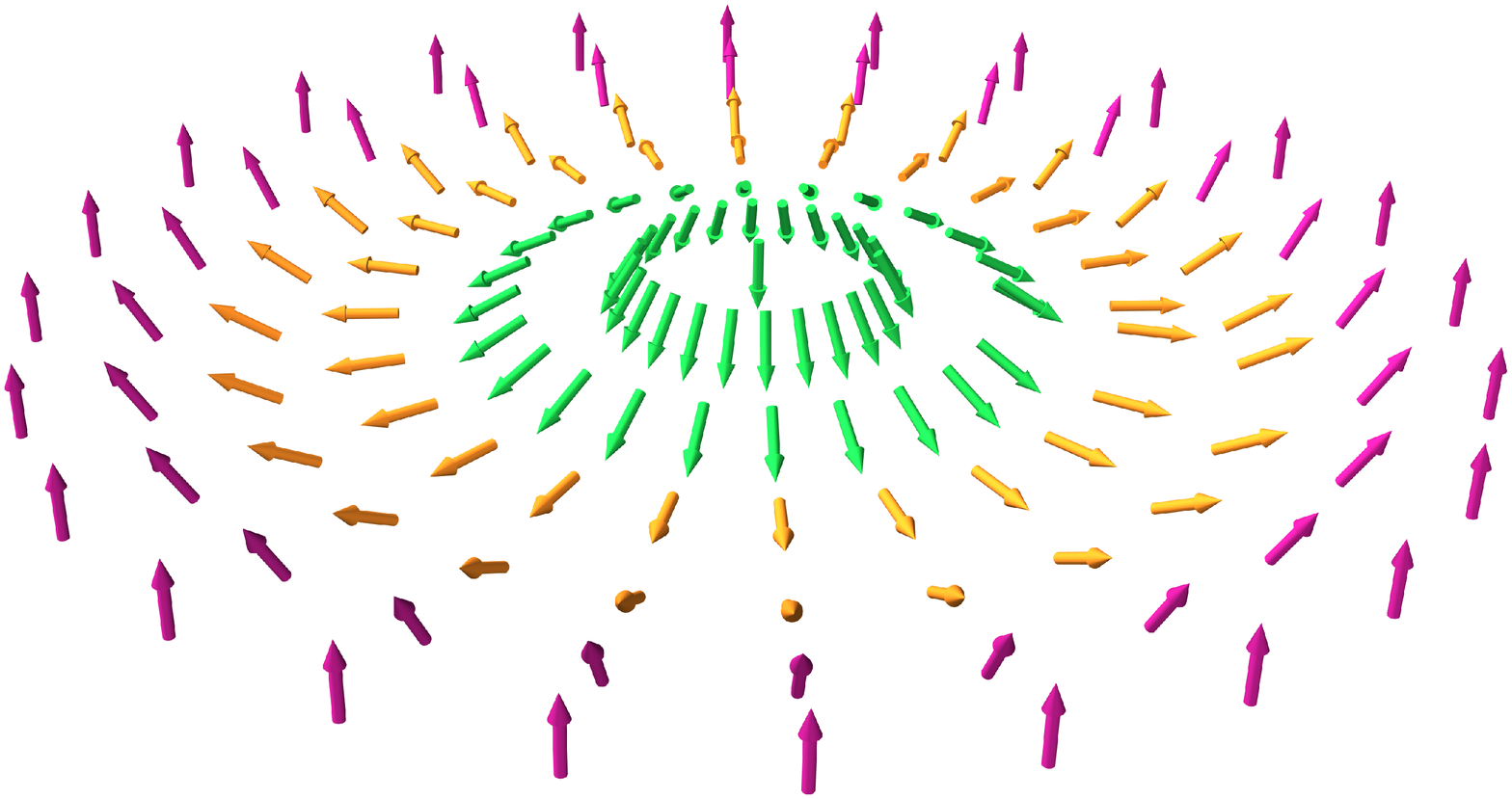}
  \caption{}
  \label{fig:skyrmion}
\end{subfigure}%
\begin{subfigure}{.5\textwidth}
  \centering
  \includegraphics[width=.55\linewidth]{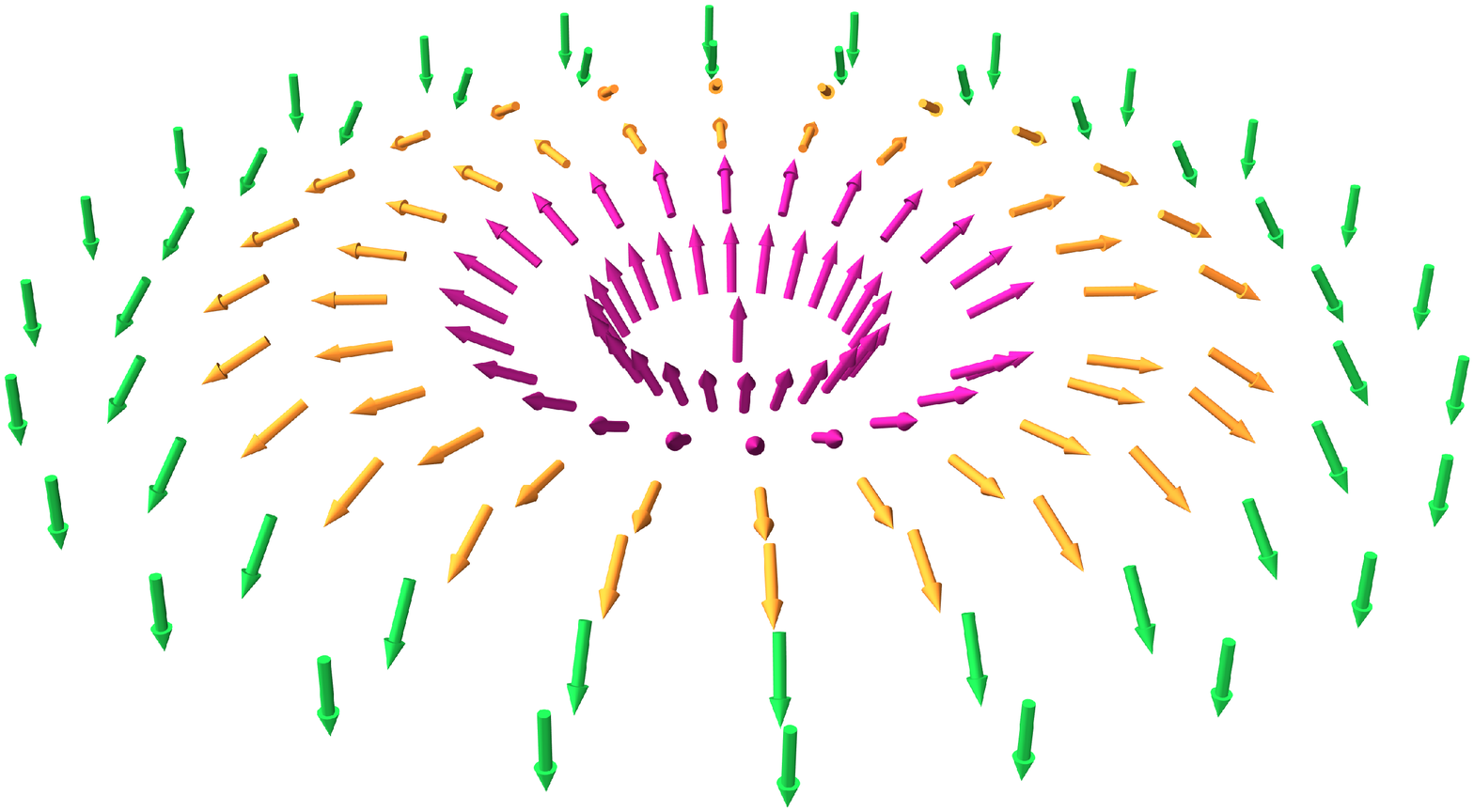}
  \caption{}
  \label{fig:anti}
\end{subfigure}
\begin{subfigure}{.5\textwidth}
  \centering
  \includegraphics[width=.55\linewidth]{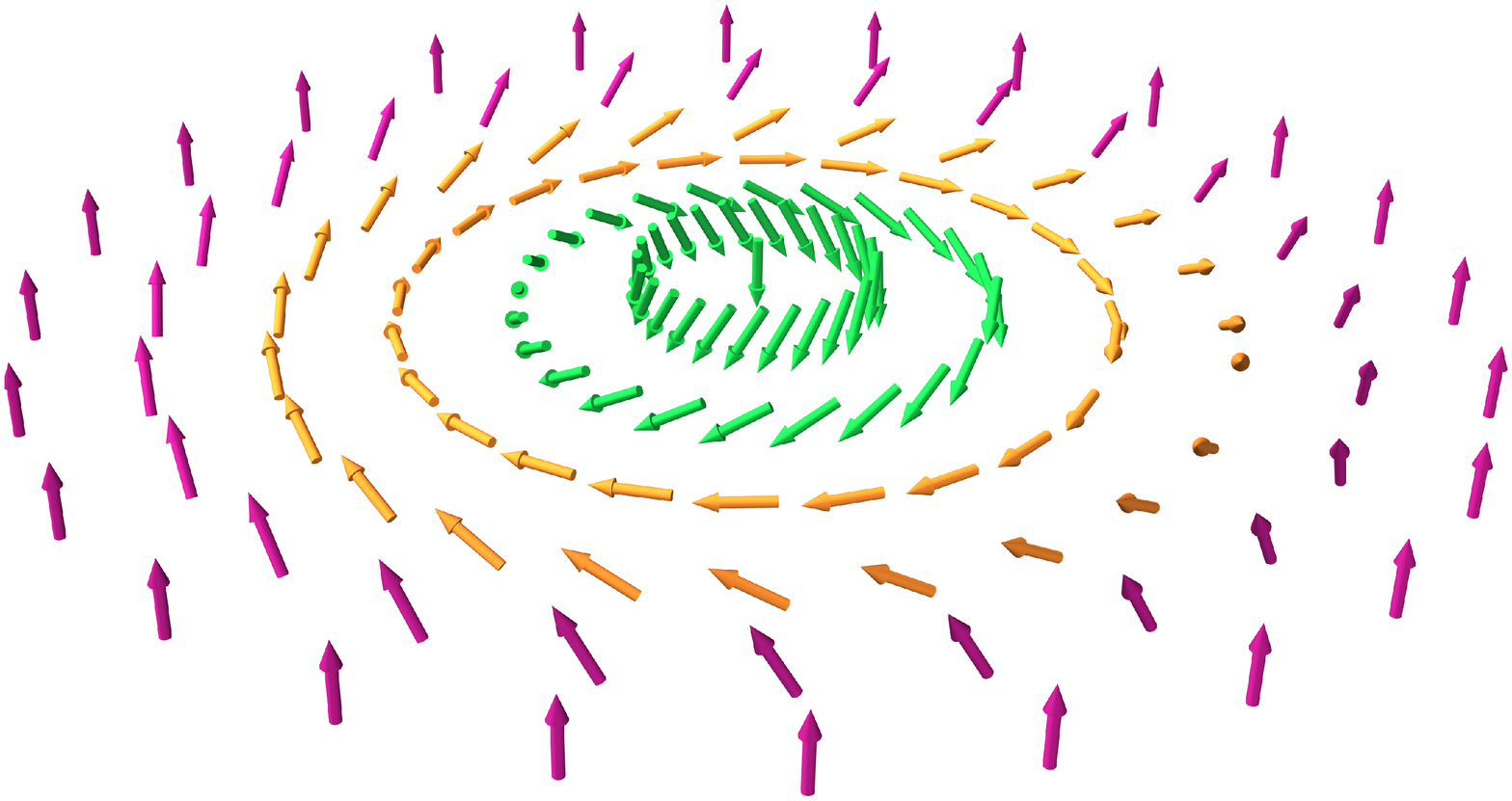}
  \caption{}
  \label{fig:anti}
\end{subfigure}%
\begin{subfigure}{.5\textwidth}
  \centering
  \includegraphics[width=.55\linewidth]{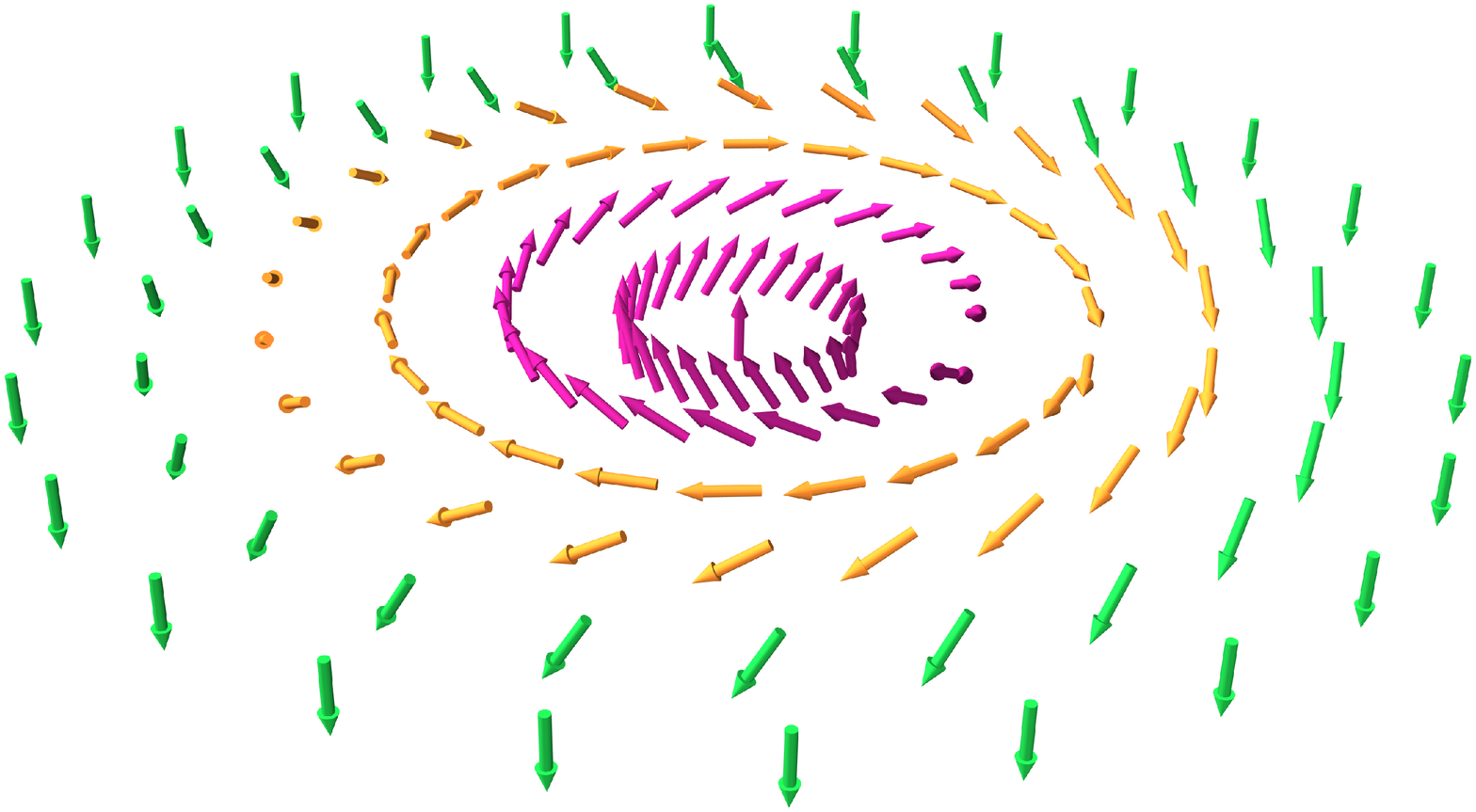}
  \caption{}
  \label{fig:anti}
\end{subfigure}
\caption{Different skyrmion magnetization textures. (a):hedgehog skyrmion, (b):hedgehog anti-skyrmion, (c):vortex skyrmion and (d): vortex anti-skyrmion. The spins at the skyrmion or anti-skyrmion center and at the skyrmion or anti-skyrmion boundary point out of plane. Inverting the chirality (helicity) of the structure is equivalent to inverting the direction only of the in-plane components of the skyrmion texture.}
\label{fig:skyrmion_structure}
\end{figure*}

Materials exhibiting the topological insulating phase, topological insulators (TIs) \cite{ti_rev,bise_ti}, have attracted a lot of attention recently, since they are promising candidates for next-generation spintronic devices due to the distinct properties of their surface states.  \cite{ti_for_mem,ti_for_mem2}. To realize such devices, it is imperative to study the interaction of TIs with other materials. During the last years, studies have been carried out regarding the interplay of TI surface state with magnetic materials, such as ferromagnets (FM)\cite{ti_for_mem,ti_mem2}. These FM can either be prepared to have a uniform magnetization e.g. magnetized along their easy axis, or support structures of varying magnetization, such as domain walls and vortices \cite{dw_vortices_on_ti}. Another magnetic configuration which can serve as a promising candidate for spintronic devices is the skyrmion \cite{skyrmion1,skyrmion2,skyrmion3}. For skyrmions to occur, strong Dzyaloshinskii-Moriya (DM) interactions and low inversion symmetry are needed. Their whirling magnetic texture has a vortex-like or a hedgehog-like shape and is characterized by the skyrmion topological number $N_{Sk}$. A schematic description of the different skyrmion magnetization textures is given in Fig. \ref{fig:skyrmion_structure}.\par
\begin{figure*}
 \includegraphics[width=\linewidth]{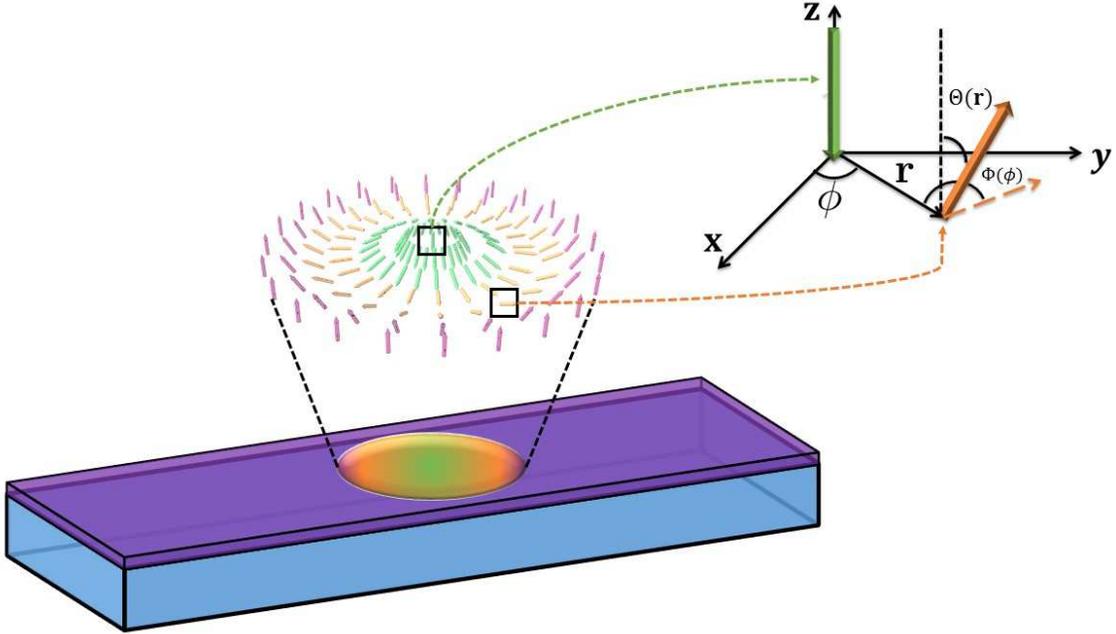}
 \caption{TI (blue) - FM (purple) bilayer with a hedgehog skyrmion (multicolored circular region) created in the thin film FM. Two particular spins have been isolated in order to explain the mathematical description of the skyrmion (eq. \eqref{eq:sk_texture}). For the shown configuration, the FM magnetization is the same as that of the skyrmion boundary (along $+\hat{\mathbf{z}}$).}
 \label{fig:FMTI}
\end{figure*}

The topological protection and small size of such whirling magnetic configurations indicate possibilities for robust and power efficient, non-volatile magnetic memory devices\cite{skyrmion3,skyrmion_mem2,skyrmion_mem3}. Therefore, the study of  their interaction with the TI surface state is important for realization of next-generation spintronic devices. Recently, a study has been published, regarding the interaction of a vortex-like skyrmion and the surface state of the 3D TI Bi$_2$Se$_3$\cite{skyrmionti}. In this paper we investigate and compare the interaction of both hedgehog and vortex skyrmions with the TI with the aim of studying the bound states that can occur on the TI surface due to the skyrmion or anti-skyrmion presence. The localized electrons in these bound states might serve as a means for electrical skyrmion detection in the future. Currently, we do not focus on how the skyrmion can be created on the FM surface in the first place. In a recent work\cite{skyrmion_logic_gate}, it is shown that conversion between different skyrmion types is possible by making material parameters or background magnetization space-dependent. Therefore, for our purpose, it suffices to assume that once we have a skyrmion magnetization texture, we can obtain also the other textures that we treat in this work. The TI/FM system we are going to study is shown in Fig. \ref{fig:FMTI}. The electrons on the surface state feel the skyrmion presence through an exchange interaction of strength $\Delta_S$ between their spins and the skyrmion magnetization texture. Consequently, an additional interaction term appears\cite{skyrmionti,Mondal2010} in the low-energy effective surface state Hamiltonian of the TI \eqref{eq:ss_hamiltonian}:
\begin{equation}
\label{eq:ss_hamiltonian}
\hat{H}=\nu_F \Big(\mathbf{p}\times\boldsymbol{\sigma}\Big)\cdot\hat{\mathbf{z}} -\Delta_S\mathbf{m}\cdot\boldsymbol{\sigma}
\end{equation}

Both the vortex and hedgehog skyrmions can induce bound states on the surface of the TI with energies $\abs{E}<\Delta_S$. However, the interactions that lead to these bound states as well as the bound state properties are different in each case. This is due to the skyrmion texture in each case and is justified in the following sections of the paper. Consequently, suitable mathematical descriptions have to be used when studying each of the vortex and hedgehog skyrmion interaction with the suface state of the TI.

This work is organized as follows. In section \ref{sec:section2} we use the variational principle to derive a numerical model for the skyrmion chiral angle $\Theta(r)$. In section \ref{sec:section3}, the skyrmion-TI interaction is studied for two models of $\Theta(r)$: in \ref{subsec:section3A} the step function approximation for $\Theta(r)$ is used\cite{skyrmionti} and in \ref{subsec:section3B} the model derived in section \ref{sec:section2} is applied to equation \eqref{eq:ss_hamiltonian}. In section \ref{sec:section4} the results from the models of section \ref{sec:section3} regarding the bound state energies and electron probability densities are presented and discussed. Finally in section \ref{sec:section5}, the most important conclusions of the present work are highlighted.

\section{\label{sec:section2}Skyrmion Description}
In general, we describe the skyrmion magnetization texture as a three-dimensional vector $\mathbf{m}$ given by equation \eqref{eq:sk_texture}. The components of this vector are functions of cylindrical polar coordinates. In this description, $\Theta(r)$ is the angle between the magnetization vector at each point in the skyrmion region and the $z$ axis. Moreover, $\Phi(\phi)$ is the angle between the projection of the skyrmion magnetization texture in the $x-y$ plane and the $x$ axis. In general, $\Phi(\phi)=\phi+\gamma$, where $\phi$ is the cylindrical coordinate and $\gamma$ is a phase which depends on the type of skyrmion we want to describe.

\begin{equation}
\label{eq:sk_texture}
\mathbf{m}=\Big(\cos(\Phi)\sin(\Theta), \sin(\Theta)\sin(\Phi),\cos(\Theta)\Big)
\end{equation}

More specifically, $\gamma$ has to do with the in-plane components of the skyrmion magnetization texture. For a specific value of $\gamma$ we get a corresponding skyrmion type with a defined \textit{helicity} $h$. For a hedgehog skyrmion, the value of $h$ determines whether the in-plane components are inward or outward whereas for a vortex skyrmion, the value of h defines a clockwise or anti-clockwise rotation of the in-plane components. The possible values for $\gamma$ are

\[
\gamma=\begin{cases}
0\;\;\; \text{, Hedgehog skyrmion (N\'{e}el) with $h$=+1}\\
\pi\;\;\; \text{, Hedgehog skyrmion(N\'{e}el) with $h$=-1}\\
\frac{\pi}{2}\;\;\; \text{, Vortex skyrmion (Bloch) with $h$=+1}\\
\frac{3\pi}{2}\;\;\; \text{, Vortex skyrmion (Bloch) with $h$=-1}
\end{cases}
\]

Using the description given by equation \eqref{eq:sk_texture}, the skyrmion topological number is readily obtained: 

\begin{equation}
\label{eq:skyrmion topological number}
\begin{aligned}
N_{sk}&=-\frac{1}{4\pi}\int^{\infty}_0 r\mathrm{d}r\int^{2\pi}_{0}\mathrm{d}\phi\; \mathbf{m}\cdot\left(\frac{\partial\mathbf{m}}{\partial x} \times \frac{\partial\mathbf{m}}{\partial y}\right)\\
&=-\frac{1}{2}\int^{\infty}_{0}\;\frac{\mathrm{d}\Theta}{\mathrm{d}r}\sin\left(\Theta(r)\right)=-\frac{1}{2}\left[-\cos\left(\Theta(r)\right)\right]^{r=\infty}_{r=0}\\
&=-\frac{1}{2}\left[-m_z(r)\right]^{r=\infty}_{r=0}
\end{aligned}
\end{equation}

As a result, $N_{sk}$ is determined by the boundaries we set on the magnetization at two particular points: $r=0$ and $r=\infty$. The values of the skyrmion texture at $r=0$ and $r=\infty$ are opposite. Consequently, for $m_z(0)=1$ and $m_z(\infty)=-1$ the skyrmion topologcial number $N_{sk}=-1$, whereas for $m_z(0)=-1$ and $m_z(\infty)=1$ we get $N_{sk}=1$. For $N_{sk}=1$ the magnetization texture is called a skyrmion whereas for $N_{sk}=-1$ we have an anti-skyrmion.\cite{skyrmion_logic_gate}

For a vortex skyrmion on the surface of a TI, we need to specify only the $m_z$ component of $\mathbf{m}$. This is due to the texture of the in-plane components of the vortex skyrmion. Here, we provide the justification of this statement. From equation \eqref{eq:ss_hamiltonian} we get

\begin{equation}
\begin{aligned}
\hat{H}&=\nu_F \Big(p_x\sigma_y -p_y\sigma_x\Big)-\Delta_S m_x\sigma_x -\Delta_Sm_y\sigma_y\\&-\Delta_Sm_z\sigma_z \Rightarrow\\
\hat{H}&=\Big(\nu_F p_x -\Delta_Sm_y\Big)\sigma_y -\Big(\nu_Fp_y +\Delta_Sm_x\Big)\sigma_x \\&-\Delta_Sm_z\sigma_z \Rightarrow\\
\hat{H}&=\nu_F\big( \mathbf{p}' \times\boldsymbol{\sigma}\big) -\Delta_S m_z\sigma_z
\end{aligned}
\end{equation}

with $\mathbf{p}'=p-\frac{e}{c}\mathbf{A}_S$ and $\mathbf{A}_S\sim\big(m_y, -m_x\big)$. The emergent magnetic field is given by 
\[\mathbf{B}=\nabla\times\mathbf{A}_S\] and as a result, 
\[B\sim\div \mathbf{m}_\parallel\]
where $\mathbf{m}_\parallel=(m_x,m_y)$.
Either using the mathematical description of the hedgehog and vortex skyrmions or simply by inspection in Fig.\ref{fig:skyrmion_structure}, it is evident that $\div\mathbf{m}_\parallel=0$ for the vortex skyrmion, while for the hedgehog skyrmion $\div\mathbf{m}_\parallel\neq0$. Consequently, the components $m_x$ and $m_y$ for the vortex skyrmion can be regarded as a gauge change equivalent to $\grad\phi$ which has zero curl. Since the hamiltonian \eqref{eq:ss_hamiltonian} is gauge invariant, we can always select a gauge which does not include $m_x$ and $m_y$ without any consequences. On the contrary, for a hedgehog skyrmion, $\div\mathbf{m}_\parallel$ is either positive or negative, depending on the skyrmion helicity. Therefore a similar gauge change as in the vortex skyrmion case does not apply here. As a result, in the case of the hedgehog skyrmion all three components of $\mathbf{m}$ need to be included. Therefore, approximations such as hard-wall approximation for the skyrmion texture can no longer  be justified and thus, a more realistic description of $\Theta(r)$ needs to be obtained.

In order to derive a better model for $\Theta(r)$, we use the variational principle for the micromagnetic energy profile of a FM \eqref{eq:micromagnetic_energy}, where the description of the magnetization in space is given by eq. \eqref{eq:sk_texture}. As mentioned in the previous section, in this work, we are not concerned about how the skyrmion is created in the first place. We assume that we can create skyrmions and can convert between different types of skyrmion structures \cite{skyrmion_logic_gate}. Taking as a fact that there is a magnetization of the form presented in equation \eqref{eq:sk_texture},
the first term in the energy functional (eq. \eqref{eq:micromagnetic_energy}) is the exchange energy density with exchange constant $J_S$, the third term is the anisotropy energy density with anisotropy constant $K_z$ and the second term represents the energy density of the anisotropic DM interactions with strength $D_{DM}$ that stabilize the skyrmion \cite{dmi1,dmi2,dmi3,dmi4}. More specifically, for the latter, it is shown\cite{bogdanovdmi} that the corresponding energy density can be written as follows:

\begin{equation}
\label{eq:DMI energy}
\begin{aligned}
\epsilon_{DMI}&=D_1 \left(m_z\frac{\mathrm{d}m_x}{\mathrm{d}x}-m_x\frac{\mathrm{d}m_z}{\mathrm{d}x}+m_z\frac{\mathrm{d}m_y}{\mathrm{d}y}-m_y\frac{\mathrm{d}m_z}{\mathrm{d}y}\right)\\
&+D_2\left(m_z\frac{\mathrm{d}m_x}{\mathrm{d}y}-m_x\frac{\mathrm{d}m_z}{\mathrm{d}y}-m_z\frac{\mathrm{d}m_y}{\mathrm{d}x}+m_y\frac{\mathrm{d}m_z}{\mathrm{d}x}\right)\\
&+D_3\left(m_x\frac{\mathrm{d}m_y}{\mathrm{d}z}-m_y\frac{\mathrm{d}m_x}{\mathrm{d}z}\right)
\end{aligned}
\end{equation}

Using eq. \eqref{eq:sk_texture} in eq. \eqref{eq:DMI energy}, we get for the micromagnetic energy \eqref{eq:micromagnetic_energy}

\begin{equation}
\label{eq:micromagnetic_energy}
\begin{aligned}
E[\Theta]=&\iiint r\; \mathrm{d}r \mathrm{d}\phi \mathrm{d}z\text{ }\Big(\epsilon_{ex}+\epsilon_{DM}+\epsilon_{anis}\Big)\\
=&\iiint r\;\mathrm{d}r \mathrm{d}\phi \mathrm{d}z\; \Bigg\{\text{ }J_S\Bigg(\Big(\frac{\mathrm{d}\Theta}{\mathrm{d}r}\Big)^2 +\Big(\frac{\sin(\Theta)}{r}\Big)^2\Bigg)\\
+&D_{DM}\Bigg[\Big(\frac{\mathrm{d}\Theta}{\mathrm{d}r} + \frac{1}{2r} \sin(2\Theta)\Big)\left(\cos(\gamma)-\sin(\gamma)\right)\Bigg]\\&+K_z \sin(\Theta)^2\Bigg\}
\end{aligned}
\end{equation}
For a specific skyrmion type, hedgehog or vortex, changing the $\gamma$ value is equivalent to a change of the sign of the DM interaction as can be seen from eq. \eqref{eq:micromagnetic_energy}.

Our purpose is to derive a realistic behavior for the skyrmion profile. In order to do this, the variational approach is applied to eq. \eqref{eq:micromagnetic_energy} and the differential equation for the angle $\Theta(r)$ which minimizes the energy functional is readily obtained:

\begin{equation}
\label{eq:differential equation for theta}
\Theta''+\frac{1}{\rho}\Theta'+\frac{\tilde{\gamma}}{2\rho}\left(1-\cos\left(2\Theta\right)\right)-\frac{1}{2}\sin\left(2\Theta\right)\left(\frac{1}{\rho^2}+\beta\right)=0
\end{equation}
where $\tilde{\gamma}=\cos(\gamma)-\sin(\gamma)$, $\beta=\frac{K_zJ_S}{D^2_{DM}}$ and $r\to\frac{J_S}{D_{DM}}\rho$.

The function $\Theta(r)$ has to satisfy eq. \eqref{eq:differential equation for theta} with the boundary conditions at $\abs{\mathbf{r}}=0$ and $\abs{\mathbf{r}}=R_S$. These conditions are defined by the skyrmion number as follows:
\begin{equation}
  \Theta(0)=\begin{cases}
               \pi\;&N_{Sk}=1\\
               0\;&N_{Sk}=-1\\
                \end{cases}
                \label{eq:skyrmion bc}
\end{equation}
and 
\begin{equation}
  \Theta(R_S)=\begin{cases}
               0\;&N_{Sk}=1\\
               \pi\;&N_{Sk}=-1\\
                \end{cases}
                \label{eq:anti-skyrmion bc}
\end{equation}
We see immediately from eq. \eqref{eq:differential equation for theta} why a step-function for the skyrmion profile is not in general a very good approximation since in that case, $\frac{\mathrm{d\Theta}}{\mathrm{d}\rho}|_{\rho=R_S}\to\infty$. Consequently, we have to disregard such solution forms.
In general, any solution that will give only one non-zero component of the magnetization would not be valid, as for the DMI energy density (eq. \eqref{eq:DMI energy}) to be non-zero, at least two components of $\mathbf{m}$ are required to be non-zero.

\begin{figure}
  \centering
  \includegraphics[width=0.9\linewidth]{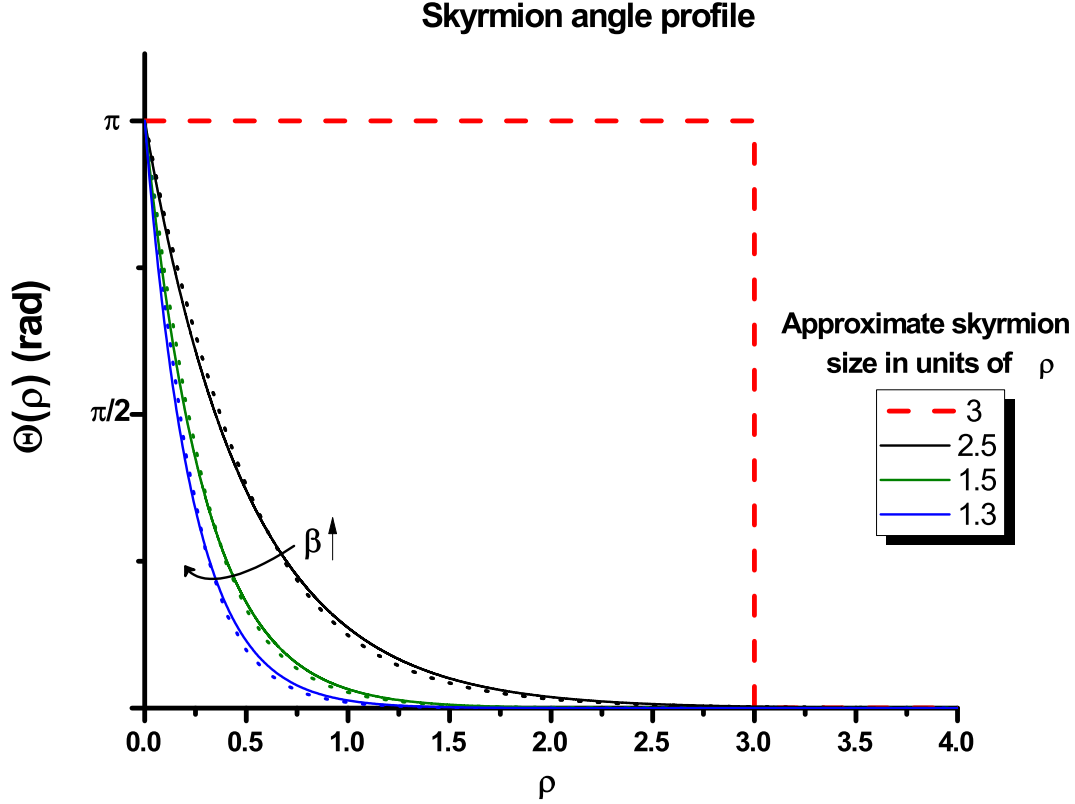}
  \caption{Angle profiles for a skyrmion texture ($N_{Sk}=1$) given by \eqref{eq:sk_texture}, for $\tilde{\gamma}=1$ and different values of $\beta$. The dotted lines correspond to the numerically computed points whereas the solid lines represent the exponentially fitted results. The step-function approximation for a specific skyrmion radius\cite{skyrmionti} is also shown in red for comparison}
  \label{fig:angle results}
\end{figure}

In Fig. \ref{fig:angle results}, the numerically computed profile (dotted line) is shown for $N_{Sk}=1$, $\tilde{\gamma}=1$ and three different values of $\beta$. From this figure, we can deduce information regarding the skyrmion size. We can identify the skyrmion radius as the first point where the angle profile has reached the boundary value and its derivative is very close to zero. A zero derivative is necessary since further from the boundary, the magnetization does not change. For example, for the black dotted curve of figure \ref{fig:angle results} we get $R_S\approx2.5$. Furthermore, the effect of increasing $\beta$ is indicated by the black arrow. For larger values of $\beta$, skyrmion size is smaller since anisotropy and exchange interactions are stronger than DMI. From our numerical calculations, we found that changing the value of $\tilde{\gamma}$ maintains the same behavior of the angle profile, though it makes it more abrupt, i.e. for the same value of $\beta$, we get a smaller effective skyrmion size. Moreover, we also found that for an anti-skyrmion, (boundary conditions \eqref{eq:anti-skyrmion bc}), the profiles can be very well approximated by $\Theta'(\rho)=-\Theta(\rho)+\pi$ where $\Theta(\rho)$ is the profile for a skyrmion. 

For computational ease when dealing with the calculation of the bound states, we do a fitting of the numerical results shown in Fig. \ref{fig:angle results}, so that we can use an analytical expression for the skyrmion or anti-skyrmion profiles. We used an exponential fitting function of the form $\pi e^{-\lambda x}$ where $\lambda$ is a constant depending on $\beta$. As shown in Fig. \ref{fig:angle results} by the solid lines, this fitting has a small error. Moreover, an exponential function for the angle $\Theta(\rho)$ also satisfies the boundary conditions for the topological number definition in eq. \eqref{eq:skyrmion topological number}. In this definition, $\left|r\right|=0$ corresponds to the skyrmion center, whereas $\left|r\right|=\infty$ corresponds to a point sufficiently far away from the skyrmion center such that we can safely say that it belongs to the FM background. However, the skyrmion magnetization profile obtained does not have a well defined radius due to the continuous transition from skyrmion region to FM background at $\left|r\right|\to\infty$.
Later in this work, when we want to calculate the bound state energies using a numerically modeled skyrmion profile,
we can set a desired skyrmion size and then calculate only the fitting parameter $\lambda$ by imposing a necessary condition on the angle $\Theta(\rho)$. Specifically, for the numerical calculations we have to introduce a cut-off value where the simulations stop. In this work, this cut-off value is given by the condition $\Theta(\rho_S)=10^{-2}\Rightarrow m_z(\rho_S)=\cos\left(\Theta(\rho_S)\right)\approx0.99$ where $\rho_S=R_S\frac{D_{DM}}{J_S}$.

\section{\label{sec:section3}Skyrmion - TI interaction }
As discussed in the previous section, for a vortex skyrmion or anti-skyrmion on the surface of a 3D TI, one needs to specify only the $m_z$ component of the magnetization texture. A crude approximation would be to consider a step-like function\cite{skyrmionti} for $m_z$ as depicted in Fig. \ref{fig:angle results} by the red dashed lines. In that case, the magnetization texture has a constant value inside the skyrmion region while after a very abrupt transition at the boundary $R_S$, it acquires the opposite value outside the skyrmion region. However, this abrupt change does not represent a very rigorous physical picture of the skyrmion structure and can lead to very different results. To illustrate this, we consider both the step function model for $m_z$ as well as the more realistic one, derived in section \ref{sec:section2}, for the calculation of the bound states. We start with the step function model for a vortex magnetization since in this case, analytical solutions for the wavefunctions can be obtained. For non-constant $m_z$ which is the case when the numerical model (eq. \eqref{eq:differential equation for theta}) is considered, no analytical solutions are possible and we have to numerically solve the problem. This is the case also when a hedgehog skyrmion is considered since for that type of skyrmion, all components of magnetization have to be present and certainly $m_z$ cannot be described by a step function.
\subsection{\label{subsec:section3A}Step function approximation}
The interaction term in this case, can be considered as an additional mass term to the Dirac-like hamiltonian \eqref{eq:ss_hamiltonian} equal to $\pm \sigma_z$. The sign of this mass term is determined by the value of the angle inside and outside the skyrmion region. In this case, eq. \eqref{eq:ss_hamiltonian} reduces to the Hamiltonian 
\begin{equation}
\label{eq:ss_hamiltonian_approx}
\hat{H}=\nu_F\Big(\mathbf{p}\times\boldsymbol\sigma\Big)\cdot\hat{\mathbf{z}} -\Delta_S m_z(\mathbf{r})\sigma_z
\end{equation}
where for $N_{Sk}=1$,
\[
  m_z(\mathbf{r})=\begin{cases}
               -1\;&r<R_S\\
               1\;&r\geq R_S\\
                \end{cases}
\]
and for $N_{Sk}=-1$,
\[
  m_z(\mathbf{r})=\begin{cases}
               1\;&r<R_S\\
               -1\;&r\geq R_S\\
                \end{cases}
\]

Due to azimuthal symmetry, the wavefunctions which are now two-component spinors can be described by a wavefunction of the following form:\cite{skyrmionti,*schmitt,*skyrmion_logic_gate}
\begin{equation}
\label{eq:spinors}
\Psi_{1,2}=e^{im\phi}\begin{pmatrix}A_{1,2}(r)\\e^{i\phi}B_{1,2}(r)\end{pmatrix}
\end{equation}
where the subscript 1 (2) denotes the region inside (outside) the skyrmion. The normalization factors have been omitted for the moment.

To search for bound states, we let $\hat{H}$ given by eq. \eqref{eq:ss_hamiltonian_approx} act on $\Psi_{1,2}$. Defining the operator
\[
p_{\pm}=e^{\pm i \phi}\Big[\frac{\partial}{\partial r} \pm \frac{i}{r}\frac{\partial}{\partial\phi}\Big]
\]
this action yields the system of equations
\begin{equation}
\label{eq:systems}
\begin{aligned}
\Bigg[\begin{pmatrix}-\Delta_S m_z\big(\mathbf{r}\big)&-\nu_F\hbar p_-\\ \nu_F\hbar p_+&\Delta_S m_z\big(\mathbf{r}\big)\end{pmatrix}\Bigg]\Psi_{1,2}=E\Psi_{1,2}
\end{aligned}
\end{equation}

Independent of the choice for $N_{Sk}$, the components $A_{1,2}(\rho)$ of the wavefunction are given by
\begin{equation}
\label{eq:comp_A1}
\begin{aligned}
A_1(\rho)=I_m (k\rho)
\end{aligned}
\end{equation}
and
\begin{equation}
\label{eq:comp_A2}
\begin{aligned}
A_2(\rho)=K_m (k\rho)
\end{aligned}
\end{equation}
where $I_m$ ($K_m$), are modified Bessel functions of the first (second) kind, $k=\sqrt{1-\zeta^2}$, $\rho=\frac{\Delta_S}{\nu_F\hbar}r$ and $\zeta=\frac{E}{\Delta_S}$. For bound states, $\abs{\zeta}<1$ and thus, $k\in\mathbb{R}$. Note that as of hereon, we use another scaled radial variable, i.e. scaled by $\frac{\Delta_S}{v_F\hbar}$ instead of $\frac{D_{DM}}{J_S}$.

However, the solution for the components $B_{1,2}(\rho)$ does depend on the value of $N_{Sk}$. Inside the skyrmion region $B_1(\rho)$ is given by
\begin{equation}
\label{eq:compB1}
B_1(\rho)= \frac{1}{\zeta-m_{1,z}}\Big(\frac{\mathrm{d}A_1}{\mathrm{d}\rho} - \frac{m}{\rho}A_1\Big)
\end{equation}
while for the region outside, $B_2(\rho)$ is given by
\begin{equation}
\label{eq:compB2}
B_2(\rho)=\frac{1}{\zeta-m_{2,z}}\Big(\frac{\mathrm{d}A_2}{\mathrm{d}\rho} -\frac{m}{\rho}A_2\Big)
\end{equation}
The dependence on $N_{Sk}$ enters through the dependence of $m_z$ on $N_{Sk}$ for both regions. The equation that yields the allowed bound state energies, is obtained by the continuity conditions of the spinor components at the skyrmion boundary $R_S$. There, $\Psi_1(\rho_S)=\Psi_2(\rho_S)$, resulting in
\begin{equation}
\label{eq:bc}
\Psi_1(\rho_S)=\Psi_2(\rho_S)\Leftrightarrow \begin{vmatrix}A_1(\rho_S) & -A_2(\rho_S)\\B_1(\rho_S) &-B_2(\rho_S)\end{vmatrix}=0
\end{equation}

\subsection{\label{subsec:section3B}Numerical model for realistic skyrmion texture}
In a more realistic case, $\Theta(\rho)$ has the behavior of the fitted profiles (solid lines in Fig. \ref{fig:angle results}) obtained from applying the variational principle to the micromagnetic energy functional. In this case, all of the components of $\mathbf{m}(\mathbf{r})$, are non zero inside the skyrmion region. For the region $r\geq R_S$ however, $m_x=m_y=0$. As a result, $\Psi_2(\rho,\phi)$ is given by eqs. \eqref{eq:comp_A2} and \eqref{eq:compB2}. For region 1, we let $\hat{H}$ of eq. \eqref{eq:ss_hamiltonian} act on $\Psi_1$. Denoting the chirality of the skyrmion by $\alpha$, and expressing the hamiltonian in terms of the in-plane and out-of-plane components of the magnetization, this operation yields a system of two differential equations:
\begin{equation}
\label{eq:system_realistic}
\begin{aligned}
A'_1 -\frac{m}{\rho}A_1 -\alpha{M_\parallel}A&=(\zeta-m_{1,z})B_1\\
B'_1 +\frac{m+1}{\rho}B + \alpha{M_\parallel}B&=-(\zeta+m_{1,z})A_1
\end{aligned}
\end{equation}
where $M_\parallel=\sin\big(\Theta(r)\big)$ and $\alpha$ denotes the chirality of the skyrmion. In that sense, it is equivalent to the phase $\gamma$ described in Sec.\ref{sec:section2}. For a hedgehog skyrmion and $\alpha=1$, the in-plane components of the skyrmion magnetization point outwards, while for $\alpha=-1$ they point inwards. For a vortex skyrmion, the two different values of $\alpha$ denote clockwise or anti-clockwise rotation of the in-plane components. Furthermore, one can also use the parameter $\alpha$ in order to describe a weighted interacton between in-plane and out-of-plane components \cite{Wakatsuki2015}. In this work however, the interaction of the components is equal both for the in-plane and out-of-plane. If we let $\alpha\to0$ or  $\left|M_\parallel\right|\to0$, then the system of equations \eqref{eq:system_realistic} describes the interaction of a vortex-like skyrmion on top of a 3D TI. In both cases, the skyrmion structure is only adequately described by the more realistic angle $\Theta(\rho)$ which was numerically derived in Sec.\ref{sec:section2}. The bound states are obtained by numerically solving the coupled equations \eqref{eq:system_realistic} together with the boundary condition given by eq. \eqref{eq:bc}.



\section{\label{sec:section4}Results and Discussion}
\subsection{\label{subsec:section4A}Bound States}
\subsubsection{\label{subsubsec:section4A1} Vortex skyrmion}
\begin{figure}
\includegraphics[width=8.5 cm]{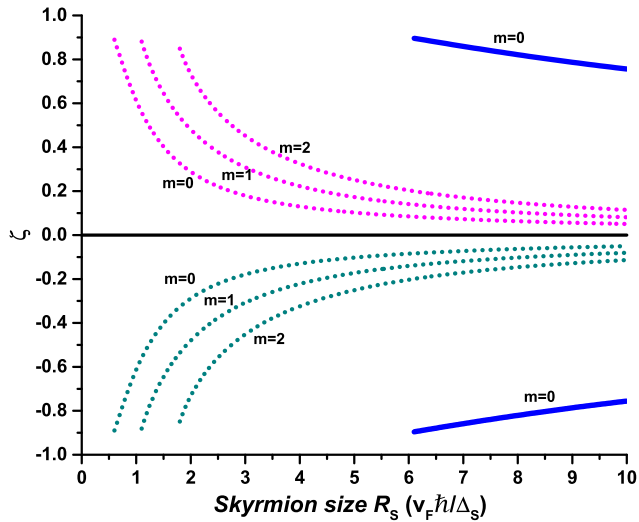}
\caption{Bound state energy ratios $\zeta$ as a function of the vortex skyrmion size for step function model (dotted lined) and numerically calculated $m_z$ (solid lines)} 
\label{fig:vortex bound states}
\end{figure}

For a vortex skyrmion structure, it suffices to use only the $m_z$ component. A crude approximation would be to use a hard-wall approximation\cite{skyrmionti} as in section \ref{subsec:section3A}. For a more realistic case, it would be better to use a more realistic profile for $m_z$ as derived in Sec. \ref{sec:section2}. In both cases, we can solve for the bound state energy ratios $\zeta=\frac{E}{\Delta_S}$. In Fig. \ref{fig:vortex bound states}, the bound state energies calculated for both of the above models are shown. The step function approximation is shown by the dotted lines whereas the numerically calculated profile is shown by the solid blue lines. For the step function profile, the cyan-colored dotted lines ($\zeta<0$) correspond to $N_{Sk}=+1$ whereas the pink-colored lines ($\zeta>0$) correspond to $N_{Sk}=-1$. The magnetic quantum number $m$ is denoted next to each curve. The results for the step-function model are in agreement with recent work \cite{skyrmionti}, where a similar study of a vortex skyrmion has been carried out. For the numerical model, the blue lines correspond to $m=0$ and $N_{Sk}=\pm 1$.

From Fig. \ref{fig:vortex bound states}, it is evident that increasing the size of the skyrmion causes the bound state energy levels to drop in absolute value with respect to $\Delta_S$. For the step function model, this can be understood from the point of view of a finite square well problem for a Dirac electron\cite{greiner}. The skyrmion effectively induces a square well potential of width $2 R_S$ and depth $\Delta_S$.
It is shown\cite{greiner} for the Dirac equation that an increased well depth results into more bound states with the lower-lying energy states being pushed even lower. If we consider the numerical model for $m_z$, then we no longer have a perfect squared-shape potential well. We can still observe some bound states, but these start appearing only for larger skyrmion sizes. This is due to the fact that the skyrmion-induced potential in this case is different than that for a step function chiral angle (eq.\eqref{eq:effective_potential}). Consequently, the minimum skyrmion size required for the appearance of bound states is strongly dependent on the used model for the skyrmion profile.

Furthermore, for the step-function model, the results of this work show that skyrmion and anti-skyrmion textures give symmetrical solutions for the bound state energies around $\abs{\zeta}=0$. More specifically, in the case of $m\geq0$, for $N_{Sk}=1$, $\zeta<0$ and for $N_{Sk}=-1$, $\zeta>0$. The sign change of $N_{Sk}$ is mathematically equivalent to a sign change of $m_z(\mathbf{r})\sigma_z$. As a result, the behavior of magnetization is exchanged in each region when there is a change in $N_{Sk}$. However, the overall behavior of the spinor components of both $\Psi_1$ and $\Psi_2$ is not altered since the boundary conditions in each region remain the same. Therefore, it is expected that we get different, and in this case opposite, bound state energies for different $N_{Sk}$. In contrast to the step-function case, when we use a numerical model for $m_z$, then bound states with both $\zeta>0$ and $\zeta<0$ appear for each value of $N_{Sk}$. As a result, the amount of bound states is doubled.

Finally, the case of $m<0$ is considered. The interaction of these electrons can be derived from equation \eqref{eq:ss_hamiltonian} by the inversion $\sigma\to-\sigma$. As a result, $\hat{H}\to-\hat{H}$, and the resulting $\zeta$ have the opposite sign compared to the case with $m>0$. Hence, for $N_{Sk}=1\;(-1)$ and $m<0$, the allowed bound state energy levels are given by the curve for $N_{Sk}=-1\;(1)$ and $m>0$ of Fig. \ref{fig:vortex bound states}. For the more realistic case of $m_z$ however, the results should remain the same, since the bound states come in pairs.

\subsubsection{\label{subsubsec:section4A2} Hedgehog skyrmion}
For a hedgehog skyrmion, equations \eqref{eq:comp_A2}, \eqref{eq:compB2}, \eqref{eq:system_realistic} and \eqref{eq:bc} are solved numerically to yield the allowed energy ratios $\zeta$. These results are shown in Fig.\ref{fig:bse_num}. This plot shows clearly that the in-plane components affect the bound states in several ways. 

First of all, the chirality of the hedgehog skyrmion can cause the bound states to disappear. For $\alpha=1$, the in-plane components are pointing outwards and no bound states exist. On the contrary, for $\alpha=-1$, the in-plane components point inwards and bound states can occur. This difference is explained by the fact that $\sign(B)=\sign(\alpha)$. Changing the sign of $\alpha$, effectively changes the sign of $\mathbf{m}_\parallel$. Thus, from the discussion in Sec.\ref{sec:section2}, $B$ also changes sign. Consequently, the non-zero emergent magnetic field in the hedgehog skyrmion case, alters the perfectly square effective potential well of the vortex skyrmion in such a way that for $B>0$, the bound states occurring in the square potential well case disappear. More specifically, from the system of equations \eqref{eq:system_realistic}, making the substitution $A\to U \rho^{-\frac{1}{2}}$, yields the Schr\"{o}dinger-like equation
\begin{equation}
\label{eq:effective_potential}
U''+\Big(\frac{1}{4}\frac{1}{\rho^2}-\frac{m^2}{\rho^2} -\alpha\frac{\mathrm{d}M_{\parallel}}{\mathrm{d}\rho} -\alpha\frac{2m+1}{\rho}M_\parallel-m^2_z\Big)U +\zeta^2U=0
\end{equation}
Thus, the effective potential that an electron on the surface of the TI senses, due to the skyrmion presence is given by the second term of eq. \eqref{eq:effective_potential} and as a result, the value of the in-plane magnetization as well as its rate of change along the radial direction and the magnetic quantum number $\abs{m}$ affect the bound state energy. Close to the skyrmion center, the potential strongly depends on $m$ and has a ${1}/{\rho^2}$ behavior. As we move further along the radial direction the in-plane magnetization components dominate.

The second effect of the in-plane components is the shift of the minimum skyrmion size required in order for bound states to exist. Comparing the lower limit of skyrmion size for the step-function model of a vortex  and a hedgehog skyrmion in Fig.\ref{fig:bse_num}, it is evident that for $m=0$, the minimum size for a hedgehog skyrmion is approximately three times larger. Contrary to the step function model, our numerical model for $m_z$ for a vortex skyrmion gives a minimum skyrmion size that is larger than that of a hedgehog skyrmion. Furthermore, the range of the bound state energies for a hedgehog texture is supressed compared to those of a vortex skyrmion. The shift in the minimum skyrmion size as well as the suppressed bound state energy range, depend on the criterion for the point $R_S$, which is used as a cut-off in the numerical calculations. For the results presented in this work, $R_S$ has been defined such that $\Theta(R_S)=10^{-2}$.

Another characteristic of the bound states induced by a hedgehog skyrmion texture, is the fact that after a critical size $R^{*}_S$, further increase of the size does not constitute decrease of $\abs{\zeta}$, as in the case of vortex skyrmion. This can be explained by considering the angle profiles shown in Fig. \ref{fig:angle results}. The angle $\Theta(\rho)$ behaves linearly until it attains the value $\frac{\pi}{2}$, that is the point where the skyrmion magnetization is completely in plane. From that point on,  the skyrmion magnetization changes less abruptly until it reaches the boundary $R_S$. Consequently, for a big enough skyrmion, there is a significant contribution from the in-plane magnetization components. This results in a more significant effect on the effective skyrmion potential (eq. \eqref{eq:effective_potential}) as already explained. The critical value $R^{*}_S$ is different for each $m$ since $V^{S}_{eff}$ depends on $m^2$. For $m=0$, it can be seen from Fig. \ref{fig:bse_num} that the critical size is $R^{*}_S\approx 5$. For higher $m$ this value is larger and is not captured by the size range presented in Fig. \ref{fig:bse_num}. It should also be noted that the skyrmion size units in Figs. \ref{fig:angle results} and \ref{fig:bse_num} are different. 
Moreover, for large skyrmion sizes, the components of the wavefunction $\Psi_2$ become very small since they are described by modified Bessel functions of the second kind. This has implications in the bound state energies as can be seen from eq. \eqref{eq:bc}. Contrary to this numerical model, in the step function model case, this decrease in the value of the wavefunctions for region 2 is balanced by the increase of the wavefunctions in region 1, which are given by modified Bessel functions of the first kind. However, with our numerical model, the wavefunction inside the skyrmion region does not behave in the  same way. Consequently, due to the different description of the magnetization texture, the form of the wavefunctions inside region 1 is altered and this affects both the appearance and the levels of the bound states.

Moreover, the absence of discontinuity at $R_S$ that is imposed by the numerical model which has been used, has lifted the distinction between the $N_{Sk}=\pm 1$ skyrmions. In the hedgehog case and in our numerical model for vortex skyrmion, both $\zeta>0$ and $\zeta<0$ are valid solutions for each $N_{Sk}$. Finally, for $m<0$ the situation is similar to the numerically modeled vortex skyrmion described in Sec. \ref{subsubsec:section4A1}.

\begin{figure}
\includegraphics[width=8.5 cm]{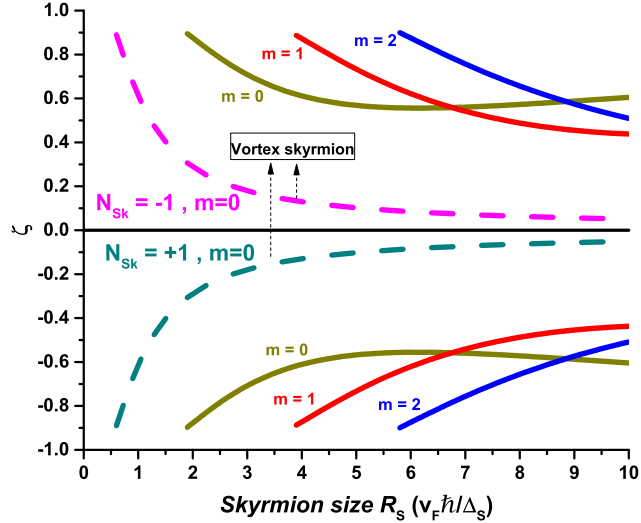}
\caption{\label{fig:bse_num} Bound state energy ratios $\zeta$ as a function of the skyrmion size. For a hedgehog skyrmion with $\alpha=-1$, both bound state solutions $\zeta>0$ and $\zeta<0$ occur for each $N_{Sk}$ (full lines). The bound state solutions for a step function model of a vortex skyrmion for $m=0$ is also shown for comparison (dashed lines). The text color corresponds to the color of the curves.}
\label{fig:bse_num}
\end{figure}

\subsection{\label{subsec:section4.2}Electronic Density}
Apart from the energy levels of these bound states, it is also useful to investigate their electronic densities. The probability density depends on the wavefunctions in each region and from equations \eqref{eq:comp_A1}-\eqref{eq:system_realistic} by extension, on the skyrmion type, $N_{Sk}$, skyrmion size, bound state energy level $\zeta$ and magnetic quantum number $m$. From the probability densities of the TI surface electrons, we can have a good estimate of the local density of states in the skyrmion region.

In Fig. \ref{fig:prob_vortex} the probability densities for three vortex skyrmions of different sizes and for $m=0$ are shown. A step-function model of $m_z$ has been used for these plots. The electrons in the bound states are localized near the skyrmion radius, $R_S$. Increasing $R_S$, lowers the probability density peak and as result the probability distribution is also broadened. Furthermore, for a vortex anti-skyrmion the same results are obtained, since in this case the analytical expressions for the wavefunctions in each region are the same as in the case of a vortex skyrmion.

\begin{figure*}
\centering
\begin{subfigure}{.5\textwidth}
  \centering
  \includegraphics[width=.8\linewidth]{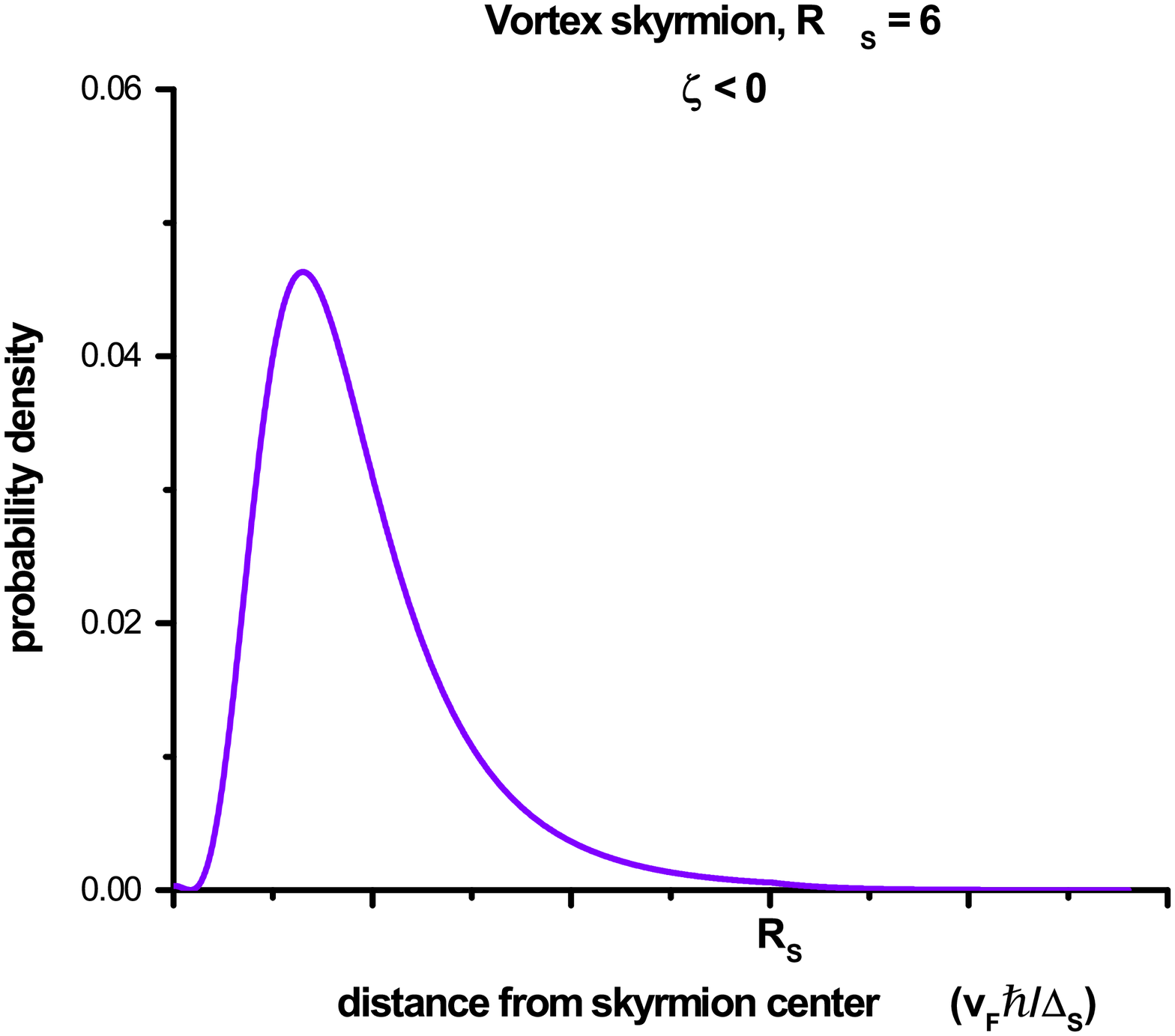}
  \caption{}
  \label{fig:vortexr6out1}
\end{subfigure}%
\begin{subfigure}{.5\textwidth}
  \centering
  \includegraphics[width=.8\linewidth]{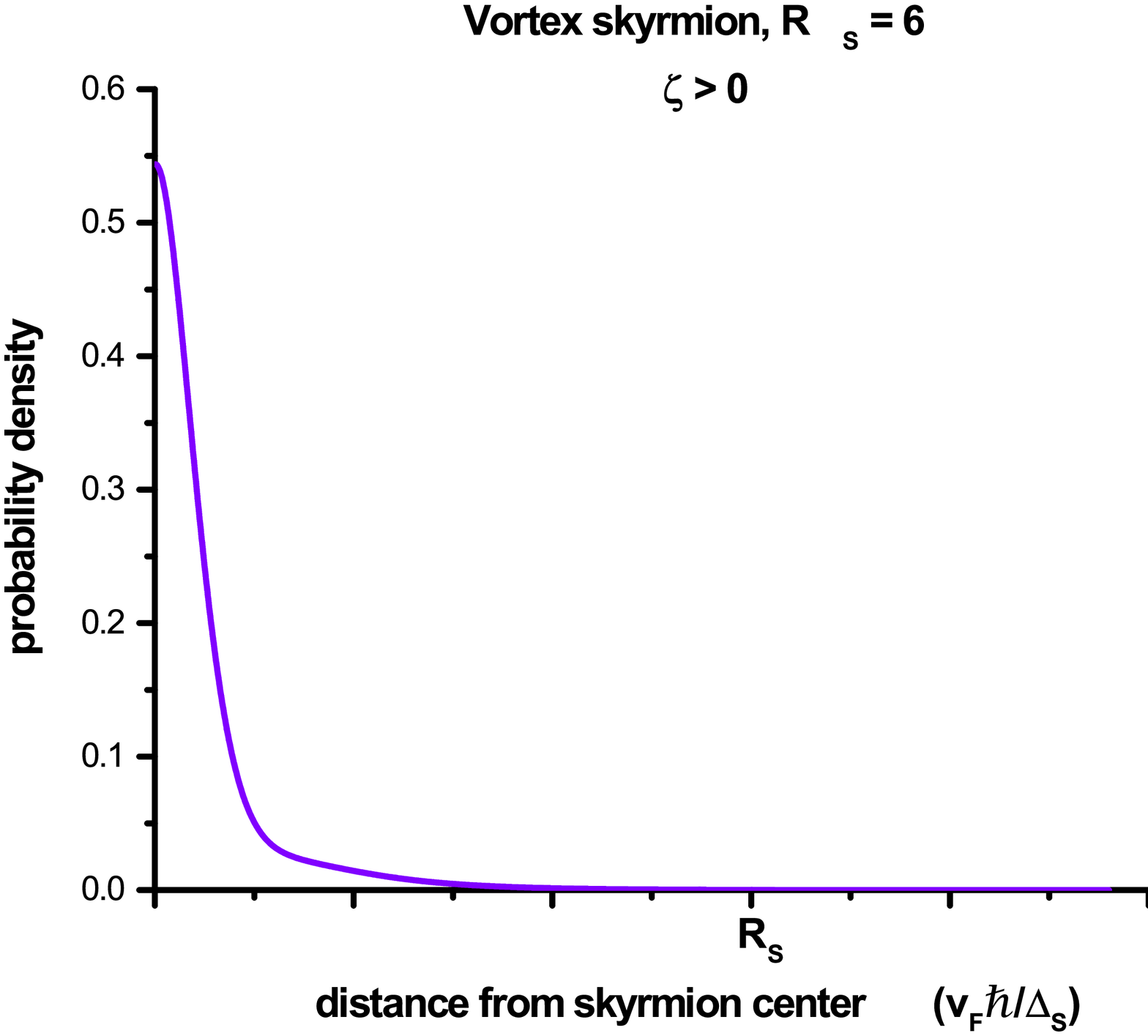}
  \caption{}
  \label{fig:vortexr6out2}
\end{subfigure}
\begin{subfigure}{.5\textwidth}
  \centering
  \includegraphics[width=.8\linewidth]{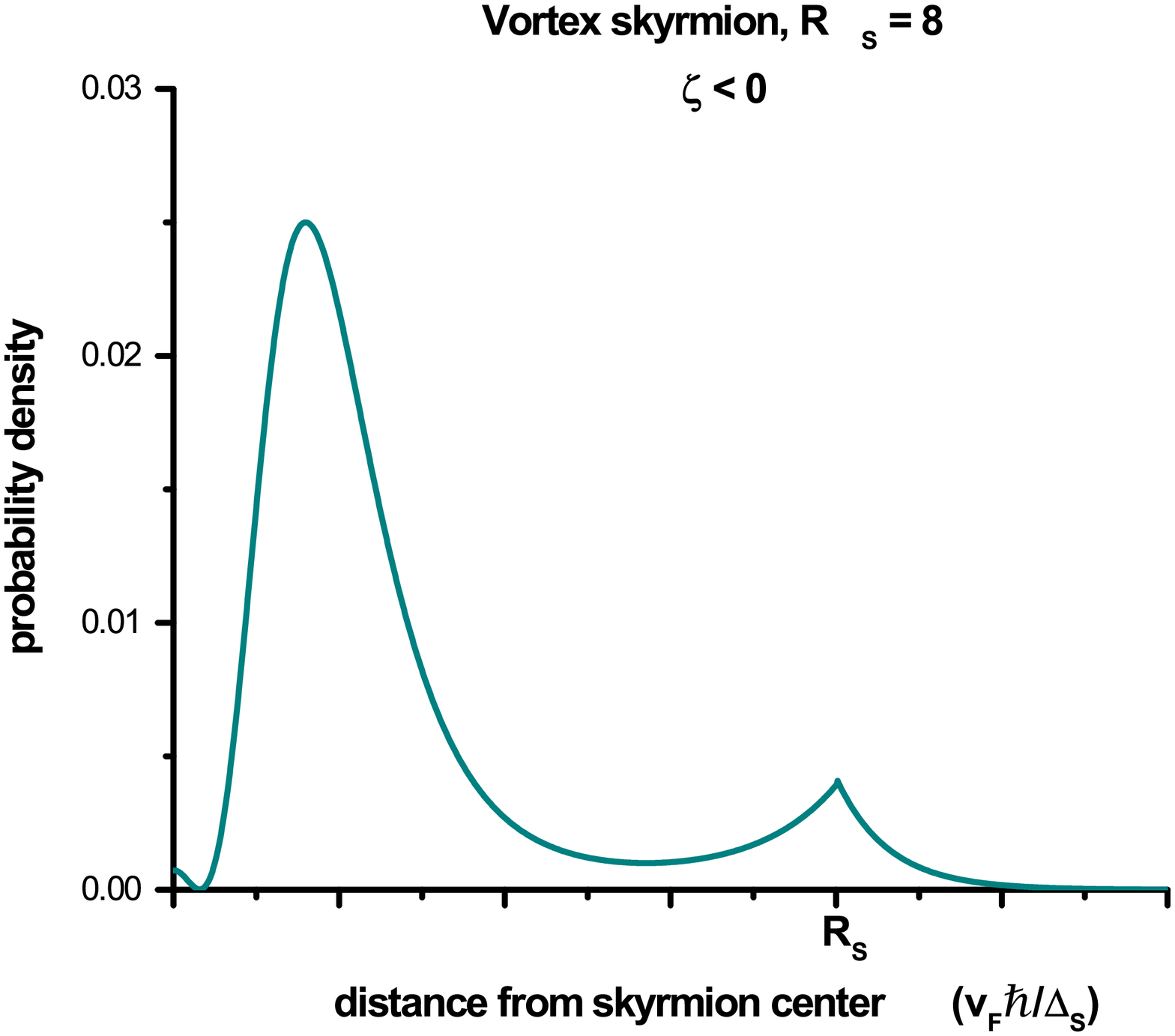}
  \caption{}
  \label{fig:vortexr8out1}
\end{subfigure}%
\begin{subfigure}{.5\textwidth}
  \centering
  \includegraphics[width=.8\linewidth]{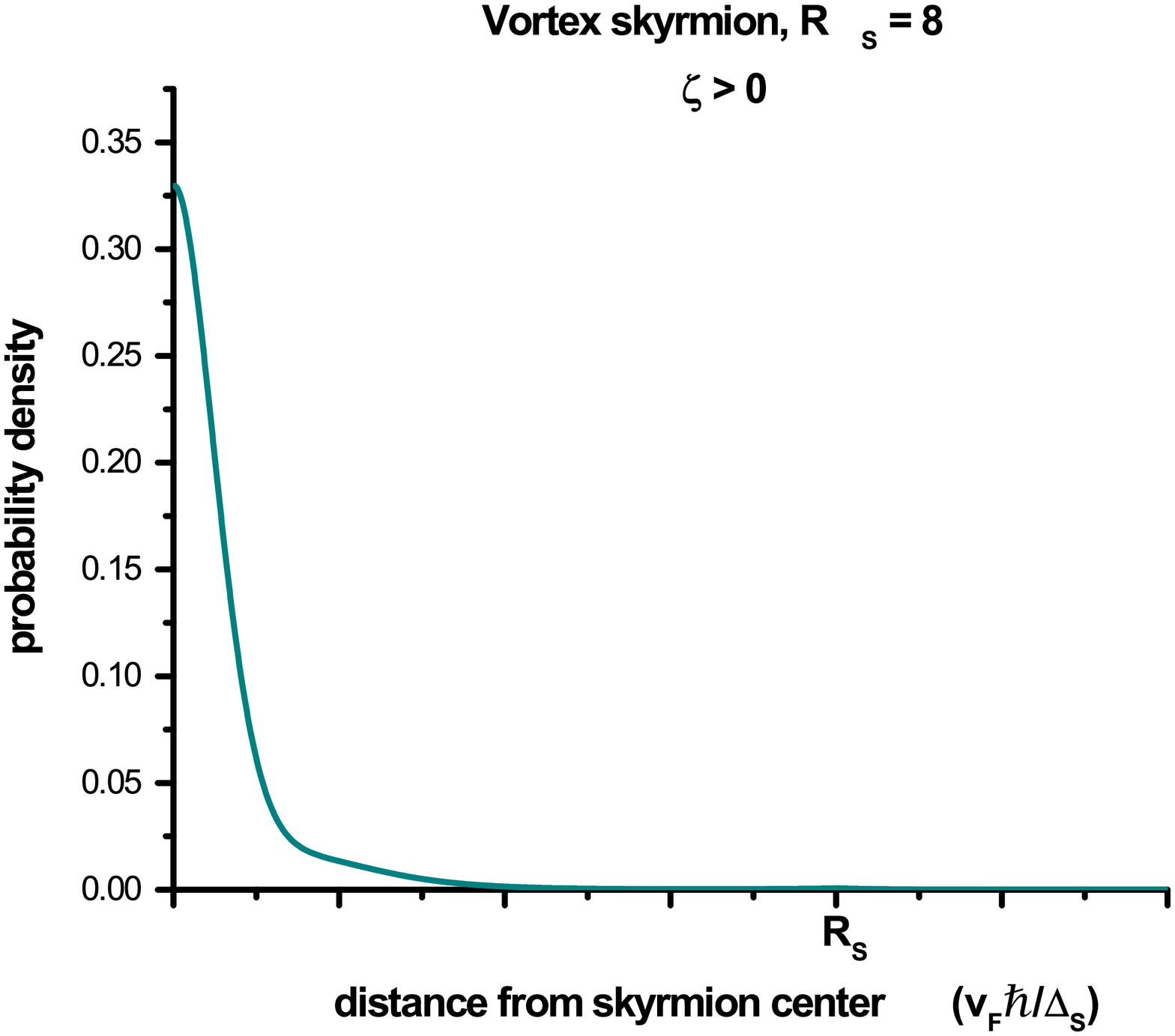}
  \caption{}
  \label{fig:vortexr8out2}
\end{subfigure}
\caption{Probability density for electrons in skyrmion-induced bound states for vortex skyrmion with sizes $R_S=6$ (figures \ref{fig:vortexr6out1} and \ref{fig:vortexr6out2}) and $R_S=8$ (figures \ref{fig:vortexr8out1} and \ref{fig:vortexr8out2}). In figures \ref{fig:vortexr6out1} and \ref{fig:vortexr8out1} the probability density is plotted for energy levels with $\zeta<0$ and in figures \ref{fig:vortexr6out2} and \ref{fig:vortexr8out2} the probability density is plotted for energy levels with $\zeta>0$. The magnetic quantum number is $m=0$ for all cases. With increasing $R_S$, the peak of the probability density decreases.}
\label{fig:electron density vortex real}
\end{figure*}

However, as is shown in  Fig.\ref{fig:electron density vortex real}, the electron density is quite different when a smooth $m_z(r)$ is considered. Moreover, the electron densities depend on the energy value of the bound state. We have shown in Figs. \ref{fig:vortex bound states} and \ref{fig:bse_num} that with a smooth skyrmion profile, the bound state solutions come in pairs. Depending on the sign of the bound state energy, the electron density is different. This is illustrated in Fig. \ref{fig:electron density vortex real}. Moreover from our simulations, when switching to an anti-skyrmion of the same size and $\zeta$, the probability density plots for $\zeta>0$ and $\zeta<0$ are interchanged. For example, for a vortex anti-skyrmion of size $R_S=6$, the probability density for $\zeta<0$ is the same as the one for a skyrmion of the same size but considering the energy level with $\zeta>0$.

\begin{figure}
\includegraphics[width=8.5 cm]{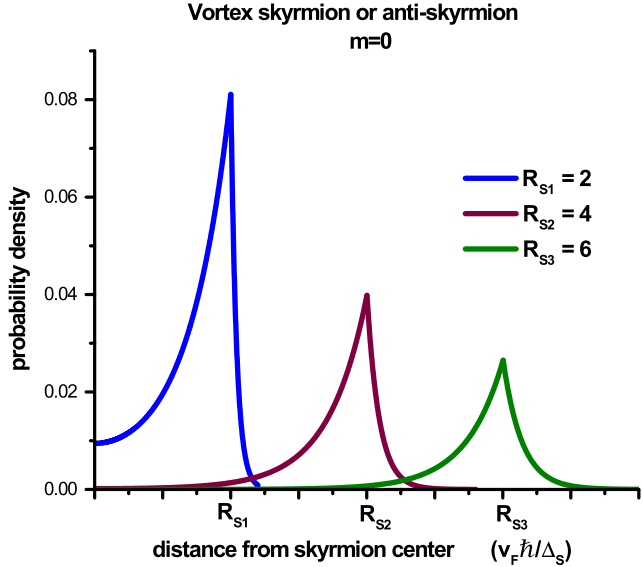}
\caption{\label{fig:prob_vortex} Probability density for a step function model of a vortex skyrmion or anti-skyrmion with size $R_S=2$ (blue line), $R_S=4$ (brown line) and $R_S=6$ (green line). }
\label{fig:prob_vortex}
\end{figure}

\begin{figure*}
\centering
\begin{subfigure}{.5\textwidth}
  \centering
   \includegraphics[width=.8\linewidth]{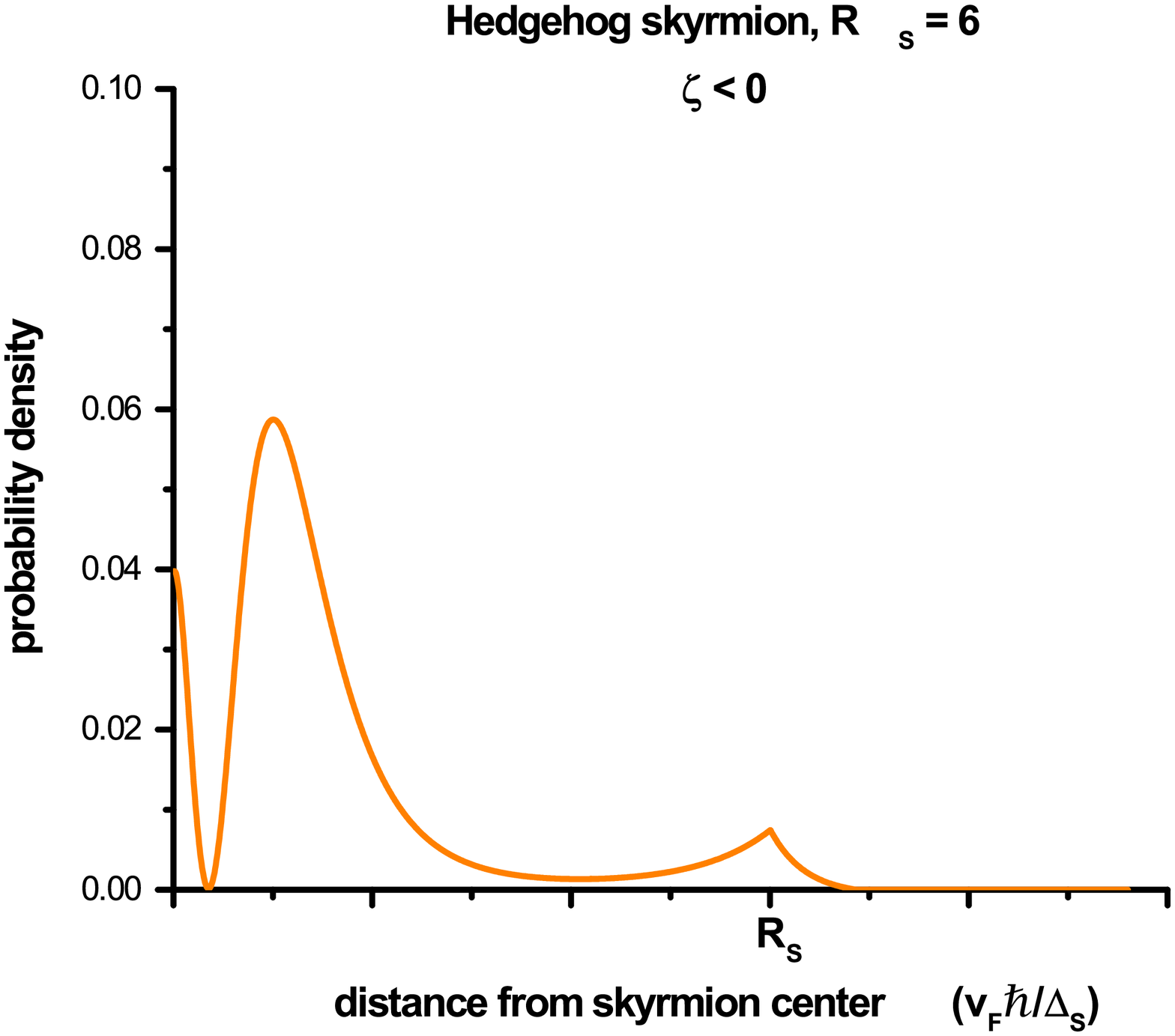}
  \caption{Hedgehog skyrmion}
  \label{fig:prob_hedge_r4sk}
\end{subfigure}%
\begin{subfigure}{.5\textwidth}
  \centering
  \includegraphics[width=.8\linewidth]{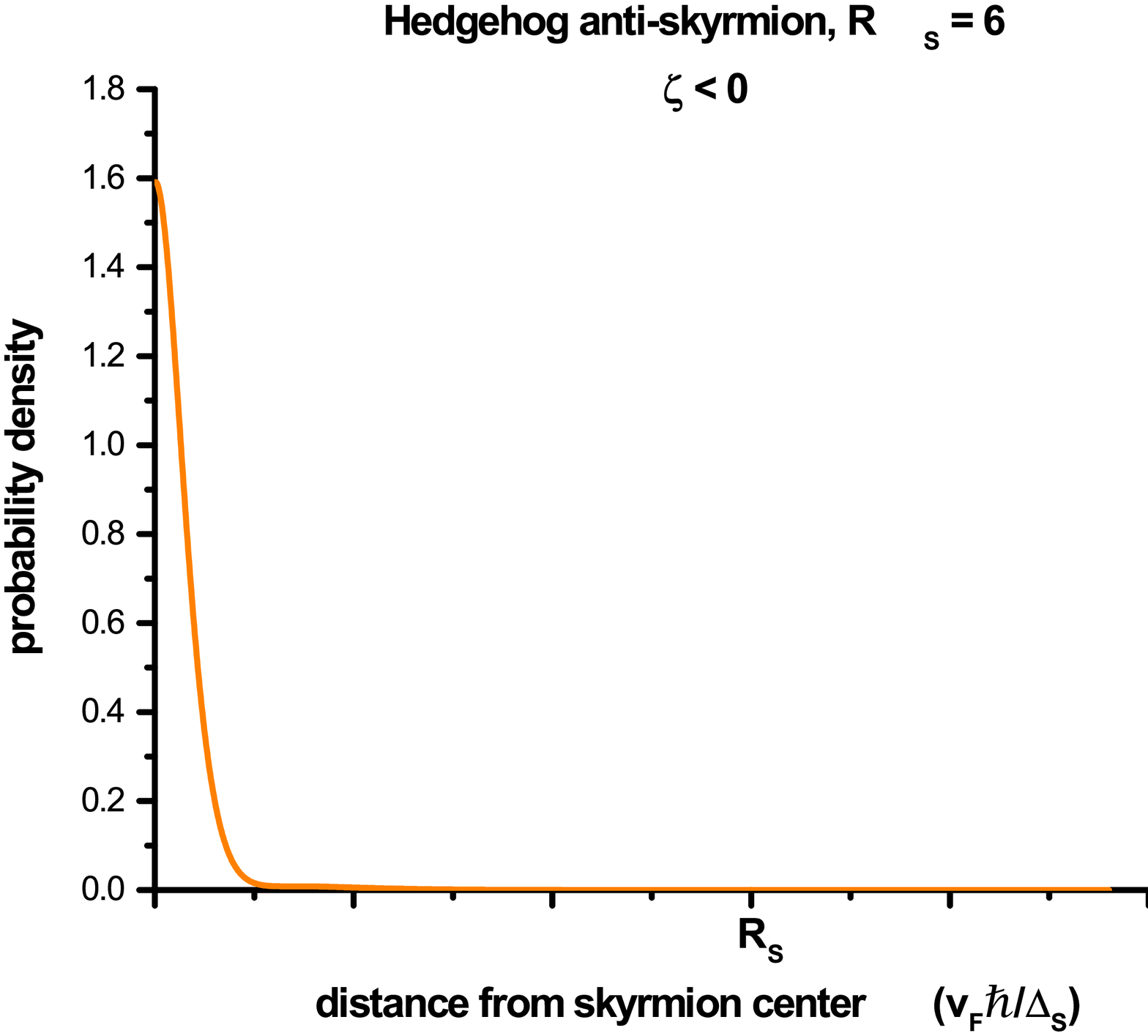}
  \caption{Hedgehog anti-skyrmion}
  \label{fig:prob_hedge_rs4ask}
\end{subfigure}
\caption{Probability densities for a hedgehog skyrmion (left) and a hedgehog anti-skyrmion (right) of size $R_S=6$ and for energy levels with $\zeta<0$ and $m=0$. Qualitatively, they are similar to the results of Fig.\ref{fig:electron density vortex real} and the quantitavie differences stem from the interaction of the in-plane magnetization components.}
\label{fig:electron density hedgehog}
\end{figure*}

For a hedgehog skyrmion, similar results can be obtained as in the case of a numerically modeled $m_z(r)$. In Fig. \ref{fig:electron density hedgehog}, the electron probability density is plotted for a skyrmion and anti-skyrmion of size $R_S=6$ for $\zeta<0$ and magnetic quantum number $m=0$. The qualitative features of the density plots for $\zeta>0$ and $\zeta<0$ are similar to those of the vortex case in Figs. \ref{fig:vortexr6out1} and \ref{fig:vortexr6out2}. However, there is an additional peak in the skyrmion case for $\zeta<0$ and the peak values of the electron density are higher. This differences are ought to the fact that for the same skyrmion size, the bound state energy levels are different for a vortex and for a hedgehog skyrmion. Furthermore, the wavefunction is different in each case, due to the fact that the in-plane magnetization components come into play in the hedgehog case. Switching to an anti-skyrmion has the same result as in the case of numerically modeled $m_z(r)$.

Putting together the results for vortex and hedgehog skyrmions when modeled with a realistic angle profile, we deduce that the electron probability density shows similar qualitative behavior for both skyrmion types (vortex or hedgehog) but distinctive features depend on skyrmion number $N_{Sk}=\pm1$, skyrmion size $R_S$ and the energy level of the bound state determined by $\zeta$. The qualitative similarity stems from the fact that the same model has been used in both cases, while the different features show the contribution of the in-plane magnetization components. More specifically, for a given skyrmion texture, the sign of the bound state will give different position and value for the probability density peak. Moreover, switching to an anti-skyrmion of the same size corresponds to interchanging the sign of $\zeta$ for a skyrmion of the same size. Finally, for each skyrmion type separately, increasing $\left|\zeta\right|$ gives larger peaks in the probabilty density.

Qualitatively speaking, the above results are similar to those of a DW texture on top of the TI, instead of a skyrmion texture. In a recent work\cite{Wakatsuki2015}, the authors have studied the interaction of N\'eel and Bloch DWs with the surface state of 3D TI and have shown that the electron density of the bound states depends on the type of DW considered as well as the interaction of the in-plane components of the magnetization texture. The results of the present work can be correlated to those of a DW if we consider the skyrmion as a loop of the DW. In fact, the same mathematical description given by eq.\eqref{eq:sk_texture}, that we have used for a skyrmion texture can be used to describe a N\'eel or Bloch DW. The only thing that changes is that the angle $\Phi$ is no longer a function of $\phi$ but it has a constant value: $\Phi=\gamma$. As in the case of DW on TIs where chiral edge modes appear along the edge of the DW, bound states can also occur for skyrmion textures, only now electrons are confined throughout the skyrmion region as shown in Figs.\ref{fig:electron density vortex real} and \ref{fig:electron density hedgehog}. Viewing the skyrmion as a DW loop, it would mean that circular (periodic) boundary conditions have to be applied to the wavefunctions of the chiral edge modes. For the skyrmion case, these have been taken into account by imposing a wavefunction of the form given by eq.\eqref{eq:spinors}.

\section{\label{sec:section5}Summary and Outlook}

In this work, we have studied the interaction between a variety of skyrmion magnetization textures and the surface state of a 3D TI. For this purpose, using a semi-classical approach we derived a numerical model of the chiral angle $\Theta$ of the skyrmion based on the variational principle, so that a more realistic behavior of the skyrmion profile could be obtained. Especially in the case of the hedgehog skyrmion on the surface of a TI, the in-plane components cannot be gauged out and thus, a step-function approximation for $m_z$ can no longer be used even as a crude approximation. The skyrmion magnetization was described by a fitting function for our numerical results and working in the macrospin approximation the interaction that was induced to the electrons of the TI was modeled by adding the extra term $\Delta_S \mathbf{m}\cdot\boldsymbol\sigma$ to the low-energy effective TI surface hamiltonian. We show that this interaction induces an effective potential given by eq. \eqref{eq:effective_potential} and thus modifies the TI surface state wavefunction. More specifically, this effective potential depends on the skyrmion type and on the way the skyrmion profile is modeled. We have shown that approximating the skyrmion profile by a step function is a very poor approximation since it does not satisfy the criteria discussed in Sec. \ref{sec:section2}. Moreover, the numerically modeled skyrmion profile gives different results for the bound state solutions.

A summary of the resulting bound states regarding the skyrmion type, the magnetic quantum number $m$ of the TI electrons and $N_{Sk}$, is shown in table \ref{table:results_bs}.
\begin{table}
\resizebox{0.4\textwidth}{!}{\begin{minipage}{0.5\textwidth}
{\renewcommand{\arraystretch}{2}
\begin{tabular}{|l||c|c|c|}
\hline
\multicolumn{4}{ |c| }{Skyrmion-TI interaction} \\ \hline
Skyrmion type & $m$  & $N_{Sk}$ &Bound States \\ \hline
\multirow{4}{*}{Vortex (step function model)} &\multirow{2}{*}{$0$}& 1 & $E^1_0 $\\
& & -1 & $E^{-1}_0$\\ \cline{2-4}
& \multirow{2}{*}{$\neq0$} &1 &$E^1_m$ , $E^{1}_{-m}$\\
& & -1 & $E^{-1}_m$, $E^{-1}_{-m}$\\ \hline
\multirow{4}{*}{Hedgehog \& numerical model for vortex} &\multirow{2}{*}{0}&1& $\pm E^{1}_0$\\ 
& & -1& $\pm E^{-1}_0$ \\ \cline{2-4}
&\multirow{2}{*}{$\neq0$} &1 & $\pm E^{1}_m$ , $\pm E^{1}_{-m}$\\ 
& & -1 & $\pm E^{-1}_m $, $\pm E^{-1}_{-m}$\\ 
\hline
\end{tabular}}\quad
\caption{Skyrmion-induced bound states on the surface of a TI. The results shown are for vortex and hedgehog skyrmions and anti-skyrmions. For the hedgehog skyrmion, $\alpha=-1$.}
\label{table:results_bs}
\end{minipage} }
\end{table}
We have shown that both skyrmion types can induce bound states. In the case of a hedgehog skyrmion, the in-plane components lead to a non-zero emergent magnetic field which has an impact on the effective potential that the TI electrons sense.
More specifically, for the hedgehog type of skyrmion, the chirality needs to be $-1$, or equivalently, the in-plane components of the skyrmion magnetization texture have to point inwards for the bound states to occur. Labeling the bound state energies with a subscript corresponding to the magnetic quantum number $m$ of the bound electrons and with a superscript corresponding to the topological invariant $N_{Sk}$, the following symmetry relations are obtained:

\begin{equation}
\begin{aligned}
\label{eq:degeneracy}
E^{N_{Sk}}_m&=-E^{N_{Sk}}_{-m}\\
E^{N_{Sk}}_m&=-E^{-N_{Sk}}_m
\end{aligned}
\end{equation}

From the data presented in Table \ref{table:results_bs} and eqs. \eqref{eq:degeneracy} we reach the conclusion that with a step function model for a vortex skyrmion, we get different energy levels for each $m$ and $N_{Sk}$. On the contrary, for a hedgehog skyrmion and for a numerically modeled vortex skyrmion, for each $m$ and $N_{Sk}$ we get $\pm E^{N_{Sk}}_m$. Consequently, the total number of bound states that we get is doubled. Moreover, the bound state energy levels $\zeta(R_S)$, depend on the skyrmion type and on the model which has been used. Therefore, the in-plane components do affect the bound states and their interaction should be taken into account. Although there is a degeneracy in the energies of the bound states implied by eqs. \eqref{eq:degeneracy}, the wavefunctions are quite different in these cases. This is demonstrated by the probability density plots of Figs. \ref{fig:prob_vortex}, \ref{fig:electron density vortex real} and \ref{fig:electron density hedgehog}. This is due to the fact that for each skyrmion type and model used, the skyrmion induced potential is different and thus the resulting wavefunction for the bound state varies. This is the main conclusion of the current work. As a result, properties such as bound state energies and probability densities will also be affected. Having knowledge of the proper skyrmion profiles, bound state energies and electronic densities is necessary for understanding and making progress in electrically detecting the skyrmion presence.


%
%

%


\bibliography{ref}

\end{document}